\documentclass{gji}
\usepackage{timet}
\usepackage{graphicx}
\usepackage{indentfirst}
\pdfoutput=1

\title[SR-Agent: An Advanced Intelligent Agent for Seismic Data Restoration]
  {SR-Agent: An Advanced Intelligent Agent for Seismic Data Restoration}

\author[Songmin Wang,etc.]
  {
  		Songmin Wang$^1$, Xintong Dong$^1$, Zilong Wang$^1$, Xinyang Wang$^2$, Zhengyi Yuan$^1$,\\   \\
  		\LARGE\upshape Wenshuo Yu$^1$\\
  $^1$ College of Instrumentation and Electrical Engineering, Jilin University, Changchun \emph{130026}, China\\
  $^2$ College of Electric Engineering, Northeast Electric Power University, Jilin \emph{132012}, China
  }

\begin{document}

\label{firstpage}

\maketitle

\begin{summary}
Field seismic data often suffer from multiple degradation mechanisms, including noise contamination, incomplete geometry, absence of low-frequency information, and insufficient spatial resolution. Therefore, some operations, such as denoising, interpolation, low-frequency extrapolation, and resolution enhancement, are often required to restore the degraded seismic data. Most existing restoration operations typically address individual degradations. When multiple degradations coexist, multiple task-specific models need to be applied sequentially, and the restoration workflow often relies on human experience to determine a suitable processing sequence. Although some all-in-one models can integrate multiple restoration operations within a unified framework, they still face a trade-off between generalization and task-specific recovery accuracy. To address this issue,  we propose a SR-Agent (Seismic Restoration Agent) containing three sequential stages: classification, restoration, and evaluation. For the first two stages, we design an intelligent classifier based on ResNet to identify the degradation types presented in low-quality seismic data, and the classification results are used as task instructions to activate the corresponding restoration models in a pool, which contains dedicated networks for different restoration tasks. For the evaluation stage, SR-Agent can dynamically schedule the activated models and explore an optimal restoration path, rather than relying on a predefined processing pipeline. Specifically, we first develop a no-reference restoration-performance evaluation measurement by combining four parameters: seismic trace coherence, energy distribution in F-K domain, spectral smoothness, and high-frequency energy ratio. Second, the evaluation measurement is applied to quantify the restoration performance of different candidate restoration paths. Finally, the path with the highest evaluation score is automatically selected and then applied to input seismic data. The results show that, when multiple degradation mechanisms coexist, SR-Agent can effectively plan an optimal restoration path, achieving accurate and intelligent restoration of degraded seismic data.
\end{summary}


\section{Introduction}

Desired seismic data are complete, noise-free, broadband, and high spatial and temporal resolution. However, due to some factors such as acquisition costs, environmental interference, and absorption and attenuation of subsurface media, the acquired seismic data are often incomplete, noisy, narrowband, and low resolution. Therefore, various restoration operations, including interpolation, denoising, low-frequency extrapolation, and resolution enhancement, are required to restore the degraded seismic data. For denoising and interpolation operations, traditional approaches include prediction filtering\cite{spi1991}, sparse representation\cite{liu2004}, and decomposition-based methods\cite{gaci2016}. Spectral extrapolation\cite{karsli2011} and phase tracking\cite{lye2015} are often used to extrapolate the missing low-frequency components of seismic data. Deconvolution\cite{rob2000} and spectral balancing\cite{zhou2016} are commonly-used resolution enhancement methods. Although these conventional restoration methods have shown good performance in specific conditions, their performance is strongly dependent on prior assumptions, such as the local linearity of seismic events, signal sparsity, wavelet estimation, sparse reflectivity assumption, seismic wavelet and attenuation characteristics, and so on. Moreover, their performance relies on manually-selected parameters and task-specific settings, limiting their adaptability to seismic data with complex and combined degradation patterns.

Over the past decade, with the rapid development of artificial intelligence, researchers have increasingly explored deep learning (DL) methods for seismic data restoration. Early studies mainly focused on individual restoration tasks, with dedicated models developed for specific types of degradation, mainly-adopted network architectures include convolutional neural network (CNN), Transformer, diffusion model, and Mamba. Yu et al.\shortcite{yu2019} proposed a CNN-based deep-learning method for seismic denoising, which learns the mapping between noisy and clean seismic data and achieves adaptive attenuation of random noise and linear noise. Dong et al.\shortcite{dong2021} proposed a CNN-based denoising method for multiarea surface seismic data, in which area-specific noise sets were constructed to match the dominant-frequency characteristics of real random noise. Mandelli et al.\shortcite{man2021} employed a U-Net-based architecture to reconstruct missing seismic traces by learning the spatial continuity of seismic events. Sun et al.\shortcite{sun2023} introduced a CycleGAN-based semi-supervised framework to extrapolate missing low-frequency components from band-limited field seismic data using unpaired field and synthetic data. Hamida et al.\shortcite{ham2022} proposed a CNN-based resolution enhancement model to reconstruct high-resolution seismic images from low-resolution data, improving the visualization and interpretation of subsurface structures. Wang et al.\shortcite{wx2025} proposed a self-supervised diffusion model (S2DM), which combines a diffusion restoration model with a variational time-spatial module to reconstruct incomplete 3D seismic data without requiring paired training data. Wei et al.\shortcite{wei2026} proposed a lightweight Mamba-based global modeling framework for 3D seismic exploration, which exploits Mamba's linear computational complexity to capture long-range dependencies and was successfully applied to 3D seismic data interpolation.

As illustrated by the above studies, most existing DL-based restoration methods are task-specific, with each model designed for a particular degradation type. However, multiple degradation phenomena often coexist in real-world seismic data, requiring multiple task-specific models and manually determined restoration sequences. This increases the complexity and computational cost of the restoration workflow and makes it highly dependent on human expertise. To alleviate these limitations, some studies have explored integrated models. Gao et al.\shortcite{gl2024} proposed a Swin Transformer convolutional residual network (SCRN) that integrates seismic denoising and interpolation within a unified framework, enabling simultaneous restoration of noisy and incomplete seismic data. Yu et al.\shortcite{yw2025} proposed a GAN-based multi-scale attention encoder-decoder network (MAE-GAN) to jointly perform seismic resolution enhancement and denoising, improving the spatial resolution and signal-to-noise ratio of post-stack seismic profiles while recovering weak geological signals. M.Cheng et al.\shortcite{cheng2025} proposed a multitask preprocessing model (MTPM) that integrates denoising, interpolation, and wavefield separation into a unified framework for DAS-VSP data, enabling these three preprocessing tasks to be performed using a single trained model. These integrated models are usually developed for a limited and predefined combination of restoration tasks, making them less flexible when the degradation types or their combinations change. 

Based on the above issues, recent studies have further explored all-in-one and foundation-model-based approaches, aiming to provide broader task coverage and stronger generalization across different seismic processing tasks and datasets. S.Cheng et al.\shortcite{cheng2025gsfm} proposed a Generative Seismic Foundation Model (GSFM), a diffusion-based all-in-one framework designed to address multiple seismic restoration tasks, including denoising, backscattered noise attenuation, interpolation, and low-frequency extrapolation. Sheng et al.\shortcite{sheng2025} proposed a Transformer-based Seismic Foundation Model (SFM) to address the limited generalization of task-specific models across different seismic processing tasks and surveys. Dong et al.\shortcite{dong2025} proposed a lightweight Seismic Processing Foundation Model (SPFM) based on the Mamba architecture to address the limited seismic data and computational resources faced by conventional foundation models, enabling efficient representation learning and transfer across multiple seismic processing tasks.

Although these all-in-one models provide greater flexibility and generalization across multiple restoration tasks, a fundamental trade-off remains between model generality and task-specific restoration performance.\cite{potlapalli2023} A unified model is expected to handle a broad range of degradation types and restoration tasks, but this broader capability may come at the cost of reduced specialization for individual tasks compared with dedicated models specifically optimized for denoising, interpolation, low-frequency extrapolation, or resolution enhancement. Conversely, independently trained task-specific models can provide stronger performance for their respective restoration tasks, but they are typically designed and optimized in isolation. In particular, existing approaches mainly focus on improving the capability of individual models to perform specific restoration tasks, without explicitly considering how these models should be intelligently coordinated and sequentially applied when multiple degradation types coexist. Therefore, an effective seismic restoration system should not only retain the task-specific advantages of dedicated models while balancing them against broader model generality, but also determine which restoration models should be invoked and in what order according to the observed degradation conditions.

This limitation is not unique to seismic data. A similar evolution has occurred in natural image restoration, where increasing complexity and diversity of real-world image degradations have gradually driven the field from task-specific models toward all-in-one restoration frameworks. Although all-in-one models improve flexibility by handling multiple degradation types within a unified network, they face similar challenges in balancing broad generalization with task-specific restoration accuracy. More recently, image restoration agents have emerged as a promising alternative paradigm \cite{ch2024,zk2025,lb2026,jx2026}. Rather than forcing a single model to perform all restoration tasks, these agents typically integrate degradation recognition, task planning, specialized model selection, model or tool invocation, and result evaluation into an adaptive workflow. By dynamically identifying the degradation characteristics of an input image and invoking appropriate specialized restoration models, an agent can retain the advantages of task-specific models while providing the flexibility required for complex degradation scenarios. This paradigm therefore provides a potential solution to the limitations of monolithic all-in-one restoration models.

\begin{figure*}
	\includegraphics[width=\textwidth]{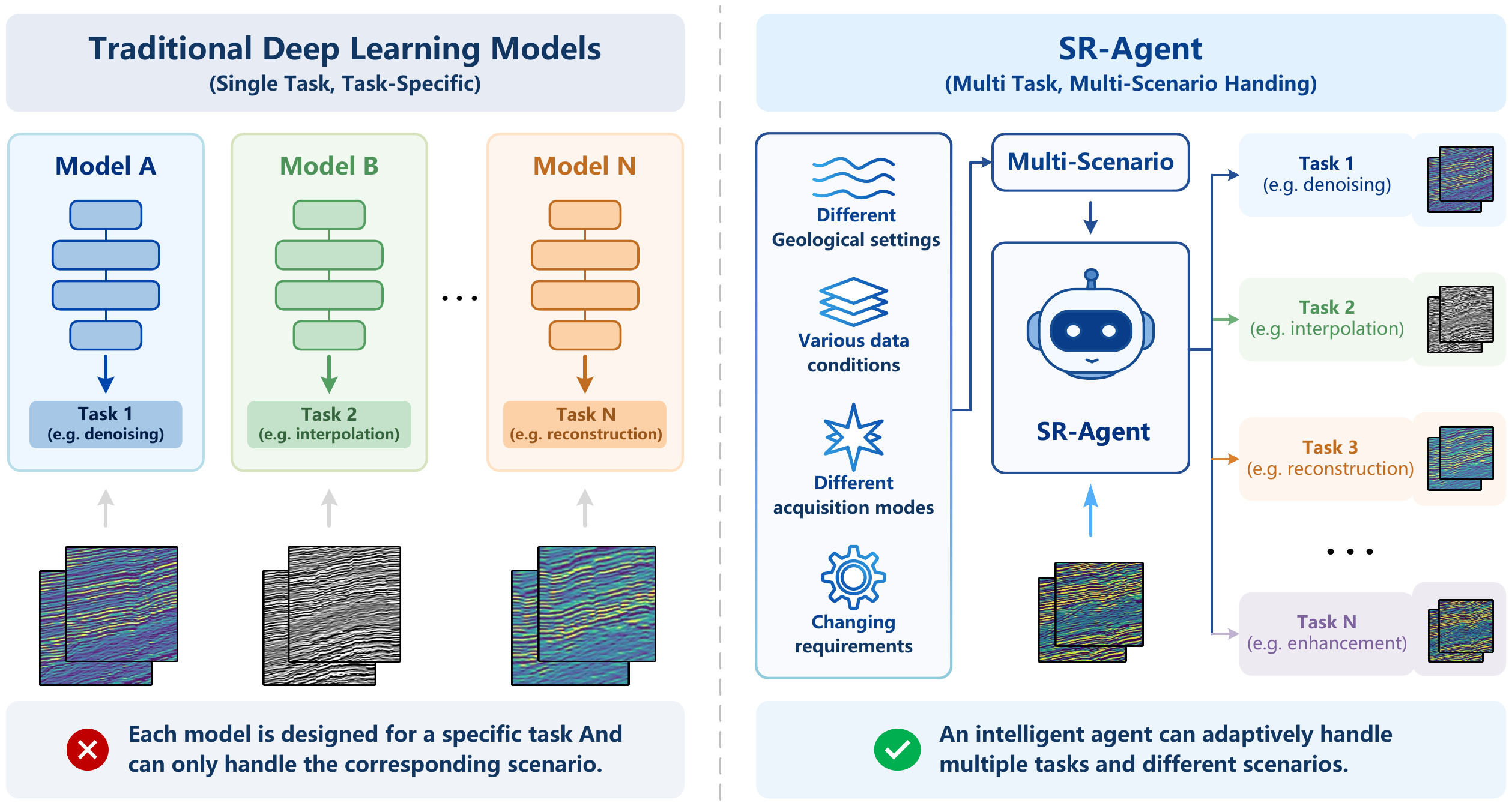} 
	\caption{Conceptual comparison of traditional task-specific models and the proposed SR-Agent for seismic data restoration.}
	\label{figure1}
\end{figure*}

Inspired by the development of image processing agents and aiming to address the aforementioned challenges, this study proposes an intelligent seismic data processing agent—SR-Agent (Seismic Restoration Agent). It can automatically identify the primary types of degradation in seismic data and independently perform processing tasks such as denoising, interpolation, low-frequency extrapolation, and resolution enhancement, enabling intelligent processing that spans data diagnosis, method selection, and result evaluation. As shown in Fig.~\ref{figure1}, compared with traditional task-specific models, SR-Agent is designed to preserve the task-specific restoration capability of dedicated models while providing greater flexibility for complex multi-degradation scenarios. Our framework provides the following functionalities:  

(1) Type identification. SR-Agent can automatically recognize the degradation type in input images and determine the corresponding restoration task.  

(2) Automated execution. Given input data, SR-Agent autonomously explores candidate restoration sequences according to the identified degradation conditions.  

(3) Optimal recovery selection. All generated results are scored using specific evaluation criteria, and the best-performing result is selected as the output. 

To achieve this, we first define multi-degradation tasks and construct a training dataset consisting of pairs of degraded images (with one or more degradation types) and their corresponding ground-truth images (used solely for evaluation). Subsequently, we train each component of the SR-Agent model so that it can autonomously detect degradation types and execute the restoration process. 

\section{METHODS}

In this section, we introduce SR-Agent, an advanced intelligent agent for seismic data restoration designed to identify the optimal combination of models and their execution sequence from a model pool to handle seismic data affected by multiple types of degradation.  
SR-Agent consists of a classification network, a model pool composed of large-scale degradation-processing models, and a computational method for evaluating restoration performance.

\subsection{Problem Definition}

We consider a comprehensive set of degradation types, denoted as $D=\{d_1, d_2, \dots, d_n\}$,
where each $d_i$ represents a specific type of seismic data degradation, such as noise, missing traces,
missing low frequencies, etc. For each degradation type $d_i$, we tailor a model library
$M_{d_i}$, comprising models $\{M_{d_i}^1, M_{d_i}^2, \dots\}$. Each model $M_{d_i}^j$ is specifically trained to mitigate the effects
of degradation $d_i$. The problem is formally defined as follows:

\vspace{0.5em}
\noindent \textbf{Input:} A degraded image $I$ subjected to various degradation types $D$. A model library
$\{M_{d_1}, M_{d_2}, \dots, M_{d_n}\}$ tailored for processing $D$. A scoring function $S$ for evaluating the image
restoration process.

\vspace{0.5em}
\noindent \textbf{Objective:} Identify the optimal model execution sequence $\sigma=\{M_{a_1}^{b_1}, M_{a_2}^{b_2}, \dots, M_{a_m}^{b_m}\}$ that
maximizes the restoration quality $S$ of the degraded image $I$, where $a_i$ denotes the degradation
type and $b_i$ represents the corresponding model. It is formulated as:
$$
\sigma^* = \arg \max_{\sigma \in \zeta(D, M)} S(I, \sigma)
$$

By addressing this issue, our aim is to determine the optimal combination of restoration
sequences and model selection, thereby improving the quality of seismic data affected by multiple degradations in real scenarios, and providing a more effective and efficient solution for complex
seismic data restoration tasks.

\subsection{Workflow Design}

\begin{figure*}
	\includegraphics[width=\textwidth]{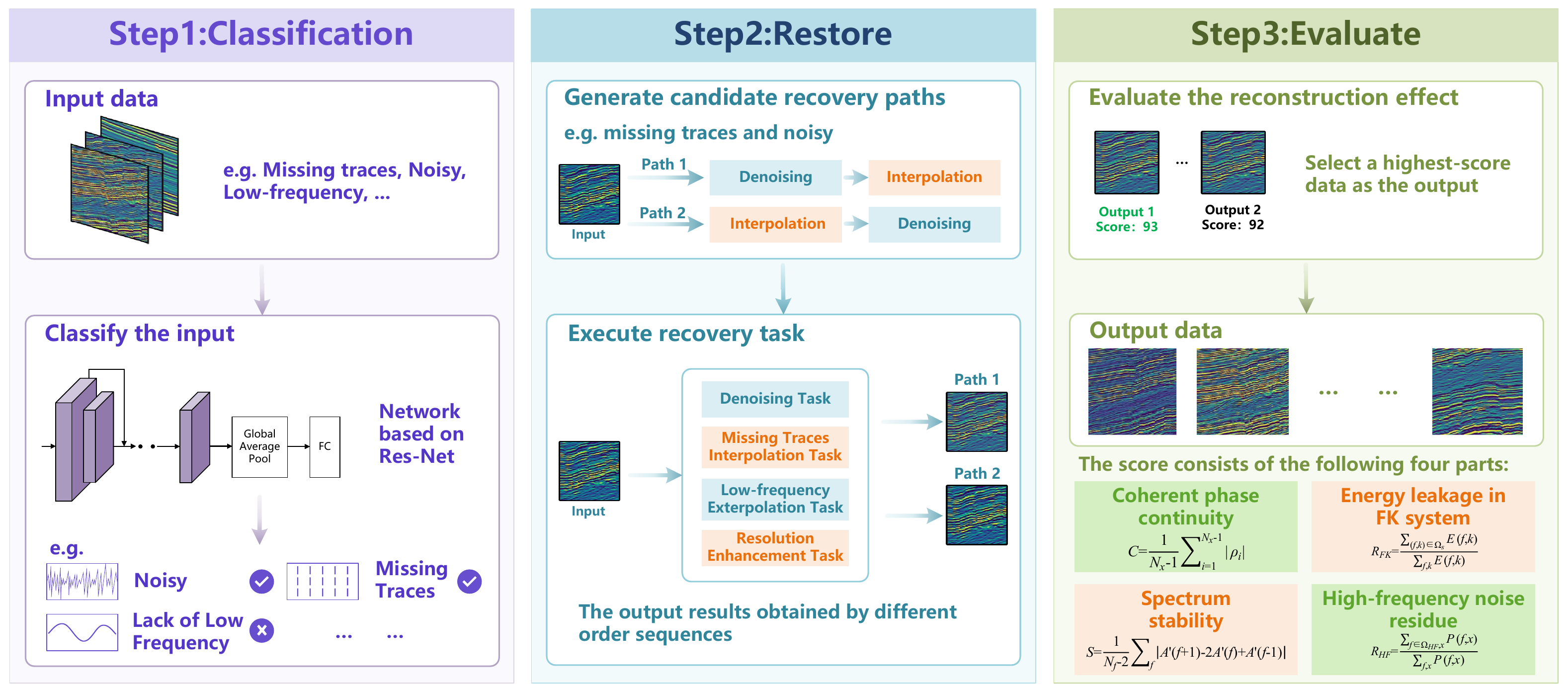} 
	\caption{The overall restoration process of SR-Agent.}
	\label{figure2}
\end{figure*}

As shown in Fig.~\ref{figure2}, SR-Agent achieves automated restoration of complex degraded seismic data through three sequential stages: classification, restoration, and evaluation.

In the classification stage (Step 1), SR-Agent first analyzes the input low-quality seismic data using a ResNet-based degradation identification network to automatically identify the degradation types, including random noise, missing traces, and low-frequency loss \cite{he2016,lj2020}. The classification network extracts degradation-related features through convolutional layers, followed by global average pooling and a fully connected layer to produce the classification results. These results indicate the degradation characteristics of the input data and determine the corresponding restoration requirements. This automated classification process eliminates the need for manual identification of degradation types and provides a basis for selecting appropriate restoration strategies.

In the restoration stage (Step 2), SR-Agent activates the corresponding specialized models from a restoration model pool according to the classification results. The model pool consists of dedicated networks for different restoration tasks, including denoising, missing-trace interpolation, low-frequency extrapolation, and resolution enhancement. Each specialized network focuses on a specific restoration task, avoiding the difficulty of forcing a single model to simultaneously learn heterogeneous degradation characteristics. For seismic data containing a single degradation type, the corresponding specialized model is directly activated to perform restoration. However, when multiple degradations coexist, different restoration operations may interact with each other, and their execution order can significantly influence the reconstruction quality. To address this issue, SR-Agent generates multiple candidate restoration paths by arranging the activated models in different execution orders. For example, for seismic data affected by both random noise and missing traces, two candidate paths can be constructed: denoising followed by interpolation, and interpolation followed by denoising. These alternative paths enable SR-Agent to explore different combinations of restoration operations rather than relying on a predefined processing pipeline.

In the evaluation stage (Step 3), SR-Agent employs a no-reference quality assessment strategy to evaluate the reconstruction results generated by different candidate restoration paths. Since clean reference data are generally unavailable in practical seismic processing scenarios, the evaluation is based on four complementary indicators: coherent phase continuity, energy leakage in the FK domain, spectrum stability, and high-frequency noise residue. These indicators characterize the continuity of seismic reflection events, the distribution of energy in the frequency-wavenumber domain, the smoothness of the frequency spectrum, and the residual high-frequency noise, respectively. By combining these four components, SR-Agent calculates an overall quality score for each candidate reconstruction result and selects the highest-scoring result as the final restored seismic data. Through this quality-based evaluation and selection mechanism, SR-Agent can automatically identify an appropriate restoration path for complex degraded seismic data without requiring manual intervention.

\subsection{Capability Acquisition}
Sec. 2.2 outlined the workflow of the proposed SR-Agent. To make this complex process both feasible and effective, the system requires several non-trivial capabilities. In the following, we describe these essential capabilities and how we equip the system to acquire them.
\subsubsection{Classification}

For an agentic system that process seismic data based on different degradation conditions, it is crucial to provide them with the ability to identify the types of seismic data degradation. Currently, in seismic data processing, there are relatively few mature models for "degradation type identification", and no mature degradation recognition network system like that in the image restoration field has been developed. Existing research mainly focuses on the restoration after degradation rather than first determining "what kind of degradation has occurred in the data". Based on this, this study proposes a seismic data degradation type classification network based on Res-Net that integrates spatial domain and frequency domain features\cite{lj2020}. This method models the seismic data degradation recognition task as a five-class problem, aiming to determine the type of degradation to which the input seismic data belongs, including "clean", "missing traces", "missing traces and noisy", "noisy", and "frequency loss" categories.

\begin{figure*}
	\includegraphics[width=\textwidth]{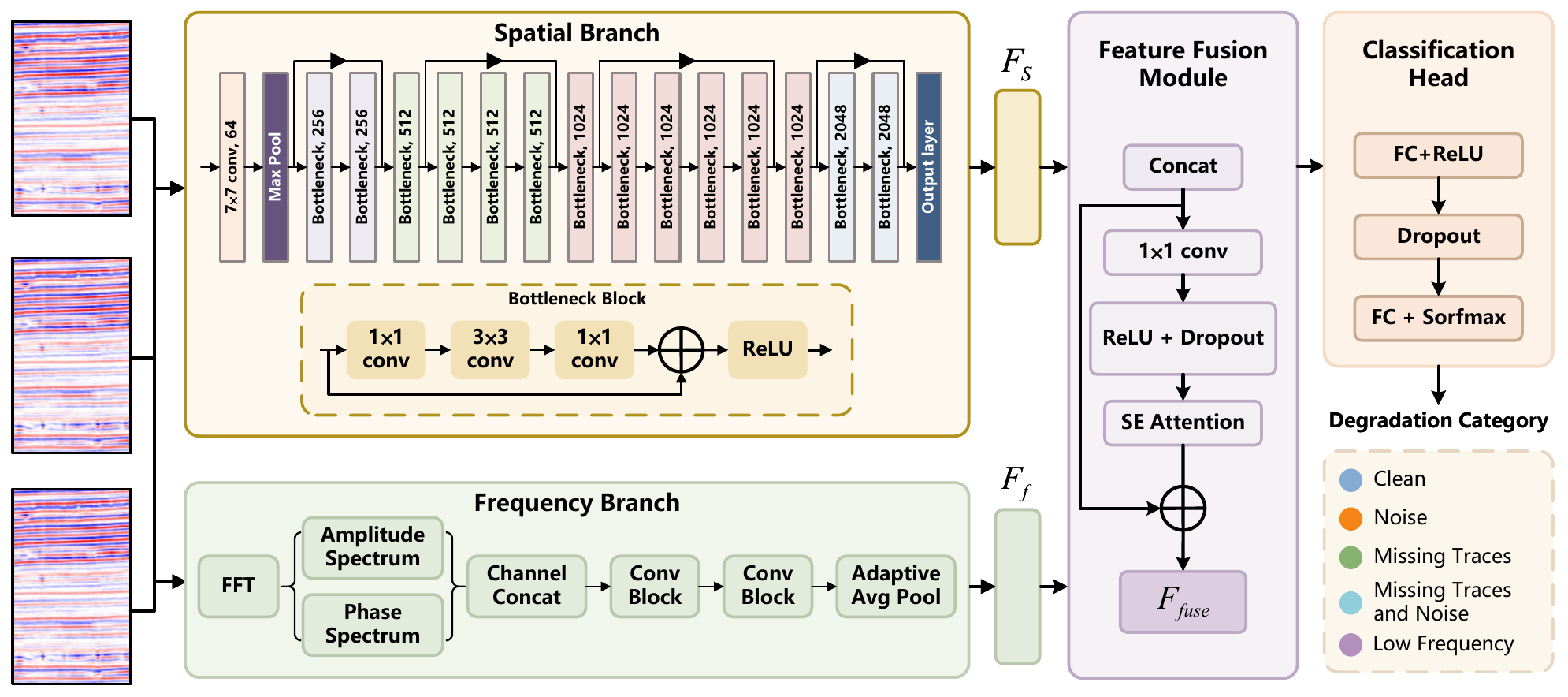} 
	\caption{The overall structure of the classification network.}
	\label{figure3}
\end{figure*}

As shown in Fig.~\ref{figure3}, the overall network structure consists of a spatial feature extraction branch, a frequency domain feature analysis branch, and a feature fusion classification module. During the training stage, a five-class cross-entropy loss function is used to optimize the model parameters, and the Adam optimizer is used for network training. The training data and validation data are input into the model through the data loading module, and the model saves the optimal weights based on the classification accuracy of the validation set after each training cycle.

In the spatial domain feature extraction aspect, the model uses a pretrained ResNet50 as the basic network structure. Since seismic data usually presents as two-dimensional single-channel data, different from the three-channel input form of natural images, this study modifies the first layer convolution of ResNet50 to a single-channel input form, enabling it to directly process seismic profile data. The network extracts spatial textures, amplitude changes, and local structural information of seismic events layer by layer through a multi-layer convolution residual structure, and finally uses deep features to complete the discrimination of degradation patterns. The main spatial representation of the model is the 2048-dimensional deep spatial features output by the last convolution module of ResNet50.

Considering that different types of seismic degradation often accompany obvious spectral changes, such as abnormal frequency energy distribution caused by noise and reduction of the spectral range due to low-frequency missing, relying solely on spatial domain information may be difficult to fully distinguish different degradation patterns. Therefore, this study further designs a frequency domain analysis module to perform two-dimensional fast Fourier transform (2D FFT) on the input seismic data, converting the seismic data from the spatial domain to the frequency domain, and extracting corresponding amplitude spectrum and phase spectrum features. Then, the amplitude information and phase information are fused in the channel dimension and further extracted as frequency domain features through convolution layers. This module can effectively capture the frequency energy distribution characteristics of seismic data and provide additional discrimination basis for degradation type recognition.

To fully utilize the spatial domain and frequency domain information, this study adopts a dual-branch feature fusion strategy. The spatial branch extracts high-level semantic features through ResNet50, and the frequency domain branch obtains frequency response features through the FFT analysis module. Since the features output by the two branches have different scales, an adaptive average pooling is first used to adjust the frequency domain features to the spatial size consistent with the deep features of ResNet, and then they are concatenated and fused in the channel dimension. The fused features are compressed and integrated through a 1×1 convolution layer and input into the global average pooling layer and the fully connected classification layer to achieve the probability prediction of the five types of seismic degradation.

Compared with traditional classification methods that only use spatial features, this model takes into account both the structural information of seismic data and the spectral change features. Among them, spatial domain features can reflect the data continuity disruption caused by missing traces and the local disturbances caused by noise, while frequency domain features can describe the frequency energy changes caused by different degradation types. Therefore, this method can more comprehensively characterize the degradation patterns of seismic data, providing reliable prior information for the subsequent selection of targeted restoration models.

\subsubsection{Task-Specific Restoration Model Pool}

Unlike conventional restoration frameworks that rely on a single model to directly map degraded seismic data to reconstructed results, SR-Agent adopts a modular restoration strategy based on a task-specific model pool. The underlying motivation is that different seismic degradation mechanisms exhibit distinct physical characteristics and require different restoration priors. Therefore, instead of forcing one unified network to simultaneously learn heterogeneous degradation patterns, SR-Agent decomposes the restoration process into several specialized restoration tasks.

The restoration model pool contains multiple expert models, where each model focuses on one specific type of seismic degradation. These models are independently optimized using degradation-specific training strategies and are regarded as callable restoration tools by SR-Agent. Once the degradation perception module determines the degradation condition of the input seismic data, the corresponding restoration models are activated and incorporated into the candidate restoration paths generated by the scheduling module. Specifically, four representative restoration capabilities are included in the current model pool: denoising, interpolation, low-frequency extrapolation, and resolution enhancement. Each capability corresponds to a dedicated restoration network designed according to the characteristics of the associated degradation mechanism.

For seismic denoising, the restoration model is designed to capture long-range dependencies between seismic samples. Since seismic noise suppression requires distinguishing coherent geological structures from irregular disturbances, a Transformer-based architecture is adopted to model global correlations within seismic sections\cite{wh2024}. For interpolation, the restoration model focuses on recovering spatial continuity from incomplete seismic observations. A U-Net-based architecture is employed to exploit multi-scale spatial features and reconstruct missing seismic traces while preserving the continuity of reflection events\cite{ht2021}. For low-frequency extrapolation, the restoration model aims to infer absent low-frequency components from available seismic information.  A U-Net-based architecture is utilized to learn the nonlinear relationship between observed frequency components and the missing low-frequency information\cite{cs2024}. For seismic resolution enhancement, the restoration model is responsible for recovering fine-scale structural details from low-resolution seismic observations. A generative reconstruction framework is introduced to enhance spatial resolution while maintaining the geological consistency of reconstructed seismic features\cite{yw2025}.

It should be noted that these restoration models are not directly connected as a fixed processing pipeline. Instead, they are treated as independent functional modules within SR-Agent. It determines whether a restoration operation is required, which model should be invoked, and when the restoration process should terminate according to the degradation recognition results and quality evaluation feedback. This modular design provides two advantages. First, each restoration model can concentrate on learning the specific characteristics of a single degradation type, avoiding the optimization conflicts caused by multi-task learning. Second, the model pool allows SR-Agent to flexibly combine different restoration operations according to the degradation condition of the input seismic data, providing the foundation for adaptive multi-degradation restoration.

\subsubsection{Evalution of Reconstruction Effect}

To enable reliable iterative decision-making in the proposed seismic data processing agent, a no-reference seismic data quality assessment module is introduced to quantitatively evaluate the restoration results during the processing procedure. Unlike conventional full-reference metrics, such as PSNR and SSIM\cite{wang2004}, which require a complete and clean reference seismic image, the proposed quality assessment module evaluates the intrinsic characteristics of the current seismic data without requiring access to the original ground-truth data. This property is particularly important for practical seismic processing, where the ideal clean seismic data are generally unavailable.Similar challenges have also
been investigated in no-reference image quality
assessment, where image quality is evaluated without
access to pristine reference images\cite{bosse2018}.

Given the restored seismic data $ ( X \in {R}^{N_x \times N_t} ) $, four complementary indicators are calculated, including seismic trace coherence, FK-domain energy distribution, spectral smoothness, and high-frequency energy ratio. These metrics characterize the spatial continuity, frequency-wavenumber structure, spectral stability, and potential high-frequency noise of the restored seismic data, respectively.

The continuity of seismic reflection events is an important indicator of seismic data quality. For two adjacent seismic traces $ x_i $ and $ x_{i+1} $, their Pearson correlation coefficient is calculated as
$$
\rho_{i}={corr}(x_{i}, x_{i+1})=\frac{\sum_{l} x_{i}[l] \, x_{i+1}[l]}{\sqrt{\sum_{l} x_{i}[l]^2 \, \sum_{l} x_{i+1}[l]^2}}
$$

The coherence score is then defined as
$$
C=\frac{1}{N_{x}-1} \sum_{i=1}^{N_{x}-1} \left| \rho_{i} \right|.
$$

A larger $ C $ indicates stronger spatial consistency between adjacent traces and therefore better continuity of seismic reflection events.

The use of trace-to-trace correlation as a measure of lateral seismic continuity is consistent with conventional coherence-based seismic interpretation methods\cite{bah1995,mar1998,yt2015}. In particular, previous studies have demonstrated that coherency measures can effectively characterize stratigraphic continuity while remaining sensitive to discontinuities and noise-related distortions.

To characterize the spatial-frequency structure of the restored seismic data, a two-dimensional Fourier transform is first applied:
$$
F(f,k) = F_{2D}\left[ X(t,x) \right]
$$
and the corresponding FK-domain energy is calculated as
$$
E(f,k) = \left| F(f,k) \right|^2
$$

A predefined valid seismic propagation region $ \Omega_{s} $ is then used to separate the main seismic energy from the remaining FK-domain components. The ratio of energy within the valid region to the total FK-domain energy is calculated as
$$
R_{FK}=\frac{\sum_{(f,k) \in \Omega_{s}} E(f,k)}{\sum_{f,k} E(f,k)}
$$

A higher $ R_{FK} $ indicates that a larger proportion of the seismic energy is concentrated in the expected FK-domain region, suggesting a more physically consistent spatial-frequency structure.

The rationale for this indicator is supported by previous FK-domain seismic processing studies, in which the dominant seismic signals tend to exhibit structured energy distributions in the frequency--wavenumber domain, while unwanted or interfering components can be identified and suppressed according to their FK-domain characteristics\cite{sch2018}. For example, FK-domain projection and filtering have been used for seismic noise removal, interpolation, and weak-phase reconstruction by exploiting the concentration and sparsity of dominant seismic energy in the FK domain.

The average amplitude spectrum along the temporal dimension is calculated as
$$
A(f)=\frac{1}{N_{x}} \sum_{i=1}^{N_{x}} \left| F_{t}\left[ x_{i}(t) \right] \right|
$$

A logarithmic transformation is subsequently applied:
$$
A'(f)=\log\left( A(f)+\epsilon \right),
$$
where $ \epsilon $ is a small numerical constant. The spectral smoothness metric is defined according to the mean absolute second-order difference:
$$
S=\frac{1}{N_{f}-2} \sum_{f} \left| A'(f+1)-2A'(f)+A'(f-1) \right| .
$$

A smaller $ S $ indicates a smoother and more stable frequency spectrum, whereas an abnormally large value may indicate spectral oscillations or artificial frequency components introduced during restoration.

The use of spectral characteristics for seismic data quality assessment is consistent with previous seismic signal-processing studies, where frequency-domain and time-frequency-domain representations are used to characterize signal attenuation, noise distribution, and signal-to-noise variations\cite{zy2014}. In particular, frequency-dependent seismic analysis has been shown to provide useful information for distinguishing effective seismic energy from noise-related components and for evaluating changes in the frequency content of processed data.

The temporal Fourier transform is used to calculate the seismic power spectrum:
$$
P(f,x)=\left| F_{t}\left[ X(t,x) \right] \right|^{2}.
$$

The high-frequency energy ratio is defined as
$$
R_{HF}=\frac{\sum_{f \in \Omega_{HF}, x} P(f,x)}{\sum_{f,x} P(f,x)},
$$
where $ \Omega_{HF} $ denotes the predefined high-frequency region. This metric is primarily used to detect abnormal high-frequency energy that may be introduced by noise or restoration artifacts. A lower value generally indicates more effective suppression of undesired high-frequency components in denoising-oriented restoration.

This definition is also consistent with previous seismic noise-analysis approaches. For example, high-frequency energy relative to the total spectral energy has been used to estimate harmonic noise energy in vibroseis seismic data, where an increased amplitude-spectrum ratio corresponds to stronger high-frequency noise contamination\cite{kur2022,zp2024}. Therefore, the high-frequency energy ratio provides a simple no-reference measure for detecting excessive high-frequency components without requiring access to a clean reference section.

To obtain a unified quality representation for agentic decision-making, the four indicators are integrated into a comprehensive quality score:
$$
Q=0.40C+0.30R_{FK}+0.15\left( 1-R_{HF} \right)+0.15\,\frac{1}{1+S}.
$$

The coherence term is assigned the largest weight because the preservation of seismic structural continuity is regarded as the primary objective of seismic restoration. The FK-domain term constrains the spatial-frequency distribution, while the high-frequency ratio and spectral smoothness terms provide complementary information for detecting residual noise and spectral artifacts.This multi-dimensional design follows the general principle of combining structural, spatial-frequency, and spectral characteristics for seismic data quality evaluation rather than relying on a single metric.

The quality assessment module is not only used to evaluate the quality of an individual restoration result, but also serves as feedback for selecting among different restoration paths. In practical seismic restoration, multiple degradation types may coexist, and different restoration operations may lead to different final results depending on their execution order. For a candidate restoration path consisting of multiple operations, the quality score Q obtained at different stages are recorded and used to characterize the overall restoration effectiveness. When multiple feasible restoration paths are available, SR-Agent preferentially selects the operation or operation sequence that produces the highest overall quality improvement. For example, if the path of denoising followed by interpolation yields a higher quality score than the path of interpolation followed by denoising, the former is selected as the preferred restoration path. In this way, the restoration order is determined adaptively according to the quality feedback rather than being manually predefined.

Importantly, the proposed quality assessment module is designed as a decision-oriented no-reference evaluation mechanism rather than a replacement for conventional full-reference quantitative evaluation. PSNR, SSIM, and other reference-based metrics remain appropriate for the final experimental evaluation when clean ground-truth data are available. In contrast, the proposed metrics are mainly used during the restoration process, where the clean reference is generally unavailable. This distinction enables the quality assessment module to provide intermediate feedback to SR-Agent while avoiding the unrealistic assumption that ground-truth seismic data are accessible during practical processing.

\section{EXPERIMENTS AND RESULTS}

\subsection{Experimental Setup}

To comprehensively evaluate the effectiveness of the proposed SR-Agent for seismic data restoration under complex degradation conditions, experiments are conducted from three complementary aspects: degradation identification, single-degradation restoration, and multi-degradation restoration. The experimental design is intended not only to evaluate the restoration performance of the task-specific models, but also to investigate whether the proposed perception–scheduling–execution–evaluation framework can effectively identify degradation conditions, select appropriate restoration strategies, and iteratively improve the quality of degraded seismic data.

\subsubsection{Data Preparation and Degradation Generation}

The experiments are conducted using 2D seismic profiles with a spatial size of 224×224. The original seismic data are regarded as high-quality reference data, from which different degradation conditions are synthetically generated. Four degradation types are considered in the restoration framework, including missing traces, random noise, low-frequency loss, and reduced spatial resolution. To improve the diversity of the training samples and prevent the models from overfitting to a fixed degradation intensity, the degradation parameters are randomly varied within predefined ranges during data generation.

For the interpolation task, a random proportion of seismic traces is removed from the input data, with the missing ratio r randomly sampled from 0.05 to 0.30. For the denoising task, additive Gaussian noise is introduced into the normalized seismic data, with the noise standard deviation $\sigma$ randomly sampled from 0.05 to 0.30. For low-frequency extrapolation, frequency components below a randomly selected cutoff frequency X of 5–7 Hz are removed from the seismic data in the frequency domain. For the resolution enhancement task, the high-quality seismic data are downsampled with a spatial scaling factor of S=2 to generate low-resolution inputs. The corresponding high-quality seismic data are retained as reconstruction targets for supervised training.

The same degradation configurations are used to construct the training samples for both the degradation classification module and the task-specific restoration models, while the specific sample organization are adjusted according to the requirements of each model. This design ensures consistency between the degradation conditions perceived by the classification module and those encountered by the restoration models during execution.

\subsubsection{Training of the Degradation Classification Network}

The degradation classification network provides the perception capability of SR-Agent and is responsible for identifying the major degradation condition of the input seismic data before restoration. During the development of this module, we find that low-resolution degradation is not suitable for direct incorporation into the ResNet-based classifier. Specifically, low-resolution seismic samples may preserve overall structural characteristics similar to those of high-quality data, making their visual and spatial features insufficiently distinguishable from the “good” category. Directly introducing low resolution as an additional classification category therefore results in considerable classification ambiguity. Accordingly, low-resolution degradation is not treated as an explicit category in the ResNet-based degradation classifier. Instead, it is assessed independently using a spectral-based resolution evaluation strategy. The dominant frequency is first calculated as the primary indicator of resolution quality, while the effective bandwidth is further introduced as a complementary criterion. This independent assessment avoids forcing the classifier to distinguish low-resolution data from visually similar high-quality data and provides a dedicated mechanism for determining whether additional resolution enhancement processing is required.

The degradation classification network is trained using 1000 Parihaka high-quality seismic samples. Multiple degradation configurations are generated from the original samples, producing a total of 5000 samples belonging to five categories: good, missing traces, missing traces with noisy, noisy, and low-frequency loss. The degradation parameters are randomly sampled within the ranges described above, allowing the classifier to learn degradation characteristics across multiple intensity levels rather than relying on a single fixed degradation condition. The resulting dataset is divided into training, validation, and test subsets using an 8:1:1 ratio. The degradation classification network is implemented using a ResNet-based architecture and trained for 50 epochs with a batch size of 16. The model achieving the highest classification accuracy on the validation set is retained as the final degradation perception model for subsequent agentic experiments. Through this training strategy, the classifier learns to provide a compact degradation representation that can be used by the subsequent scheduling module to determine the appropriate restoration operation.

\subsubsection{Training of Task-Specific Restoration Models}

After establishing the degradation perception module, the task-specific restoration models are independently trained to provide the execution capability of SR-Agent. As illustrated in Figure 5, the high-quality seismic data are first used to generate degraded samples at multiple degradation levels according to the corresponding task-specific degradation mechanisms. Each degraded sample is then paired with its original high-quality seismic data and used to train the corresponding restoration network in a supervised learning manner.

Four specialized restoration models are employed for the four restoration tasks. The specific model usage has been described in Section 2.3.3. For each restoration model, the degraded seismic data are provided as inputs, while the corresponding high-quality seismic data served as reconstruction targets. The models are optimized using the L2 loss:
$$
Loss=\frac{1}{mn} \sum_{i=1}^{m} \sum_{j=1}^{n} \left( x(i,j)-y(i,j) \right)^{2}
$$
where $ x(i,j) $ and $ y(i,j) $ denote the predicted and target seismic amplitudes at spatial location (i,j), respectively, and $ m $ and $ n $ represent the spatial dimensions of the seismic sample. The use of supervised training with high-quality seismic data as targets enables each restoration model to learn the mapping from a specific degraded state toward the corresponding high-quality seismic representation.

This task-specific training strategy is consistent with the design philosophy of SR-Agent. Rather than requiring a single network to simultaneously learn all degradation mechanisms, each specialized model is optimized for a well-defined restoration task. The trained models therefore serve as independent execution modules that can be dynamically selected by SR-Agent according to the degradation condition identified during inference.

\subsection{Evaluation of Degradation Identification}

Accurate degradation identification is the first stage of SR-Agent and directly determines the candidate restoration models considered during subsequent scheduling. Therefore, the classification performance is first evaluated independently.

The classification network is evaluated on a test set containing 100 samples for each degradation category, resulting in a total of 500 test samples. The confusion matrix is shown in Figure 6. The classification model achieved an overall accuracy of 97\%.The precision, recall, and F1-score data is shown in Table 1.

 \begin{figure}
	\includegraphics[width=\linewidth]{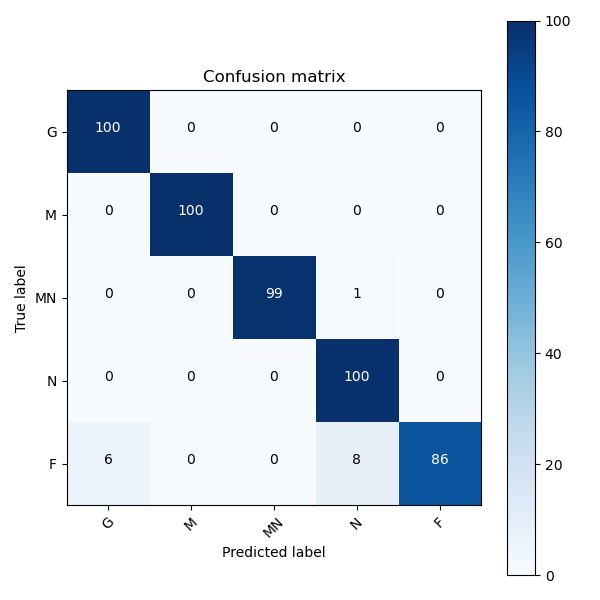} 
	\caption{Confusion matrix result of the classification network, where G represents "good", M represents "missing traces", MN represents "missing traces and noise", N represents "noise", and F represents "frequency loss".}
	\label{figure6}
\end{figure}

\begin{table}
	\centering
	\caption{The precision, recall, and F1-score data of the classification network.}
	\begin{tabular}{lccc}
		\hline
		Kind of degradation       & Precision & Recall & F1-score \\
		\hline
		Good                      & 0.94      & 1      & 0.97     \\
		Missing traces            & 1         & 1      & 1        \\
		Missing traces and noise  & 1         & 0.99   & 0.99     \\
		Noise                     & 0.92      & 1      & 0.96     \\
		Frequency loss            & 1         & 0.86   & 0.92     \\
		\hline
	\end{tabular}
\end{table}

The results indicated that the proposed spatial–frequency dual-branch architecture can effectively distinguish different degradation patterns. In particular, missing traces produced characteristic discontinuities in the spatial domain, whereas noise and frequency loss mainly altered the spectral distribution. By jointly exploiting spatial structural information and frequency-domain characteristics, the classifier could obtain complementary representations of different degradation types.

The confusion matrix further showed that the classification performance was relatively stable across the five categories. The main confusion occurred for missing low-frequency components, which were sometimes misclassified as good or noise. This could be attributed to the fact that low-frequency loss could be visually similar to good-quality seismic data, while the current FFT-based module may not always provide sufficiently distinctive frequency-domain features for this type of degradation. In addition, some low-frequency-deficient seismic images may exhibit visual characteristics similar to noise. Nevertheless, the overall classification performance provided sufficiently reliable degradation information for subsequent restoration scheduling.

These results demonstrated that degradation recognition could serve as an effective perception module for the proposed agent. Rather than directly applying a predetermined restoration sequence, SR-Agent first determined the degradation state of the input seismic data and then constructed the candidate restoration operations accordingly.

\subsection{Single-Degradation Restoration}
Before evaluating the complete agentic framework, the restoration capability of each task-specific model is independently investigated. This experiment is important because SR-Agent relies on a model pool consisting of specialized restoration networks. Therefore, the effectiveness of SR-Agent is fundamentally dependent on whether each individual model can provide reliable task-specific restoration.

 \begin{figure}
	\includegraphics[width=\linewidth]{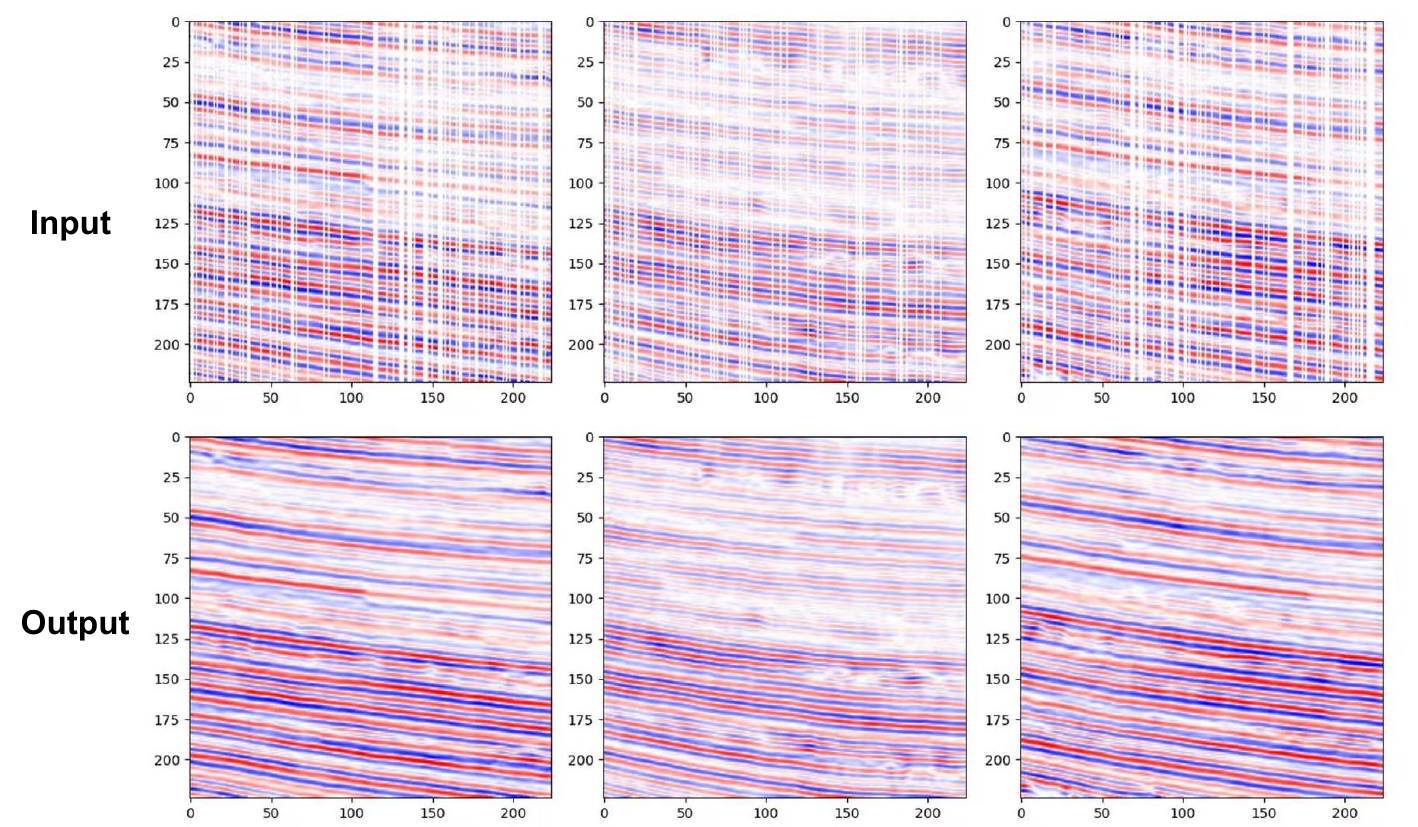} 
	\caption{The result diagram of SR-Agent when facing a single degraded seismic data with missing traces.}
	\label{figure7}
\end{figure}

Fig.~\ref{figure7} shows the restoration results for seismic data with randomly missing traces. The missing traces introduced discontinuities in the spatial direction and interrupted the continuity of seismic events. After restoration, the missing information was reconstructed based on the spatial correlations learned by the U-Net-based interpolation network. The interpolation model achieved a PSNR of 39.6698 dB and an SSIM of 0.9981. The reconstructed seismic events exhibited improved spatial continuity while maintaining the main amplitude and structural characteristics of the original data.These results confirmed that the interpolation network could serve as a specialized restoration tool for the missing-trace degradation identified by the perception module.

 \begin{figure}
	\includegraphics[width=\linewidth]{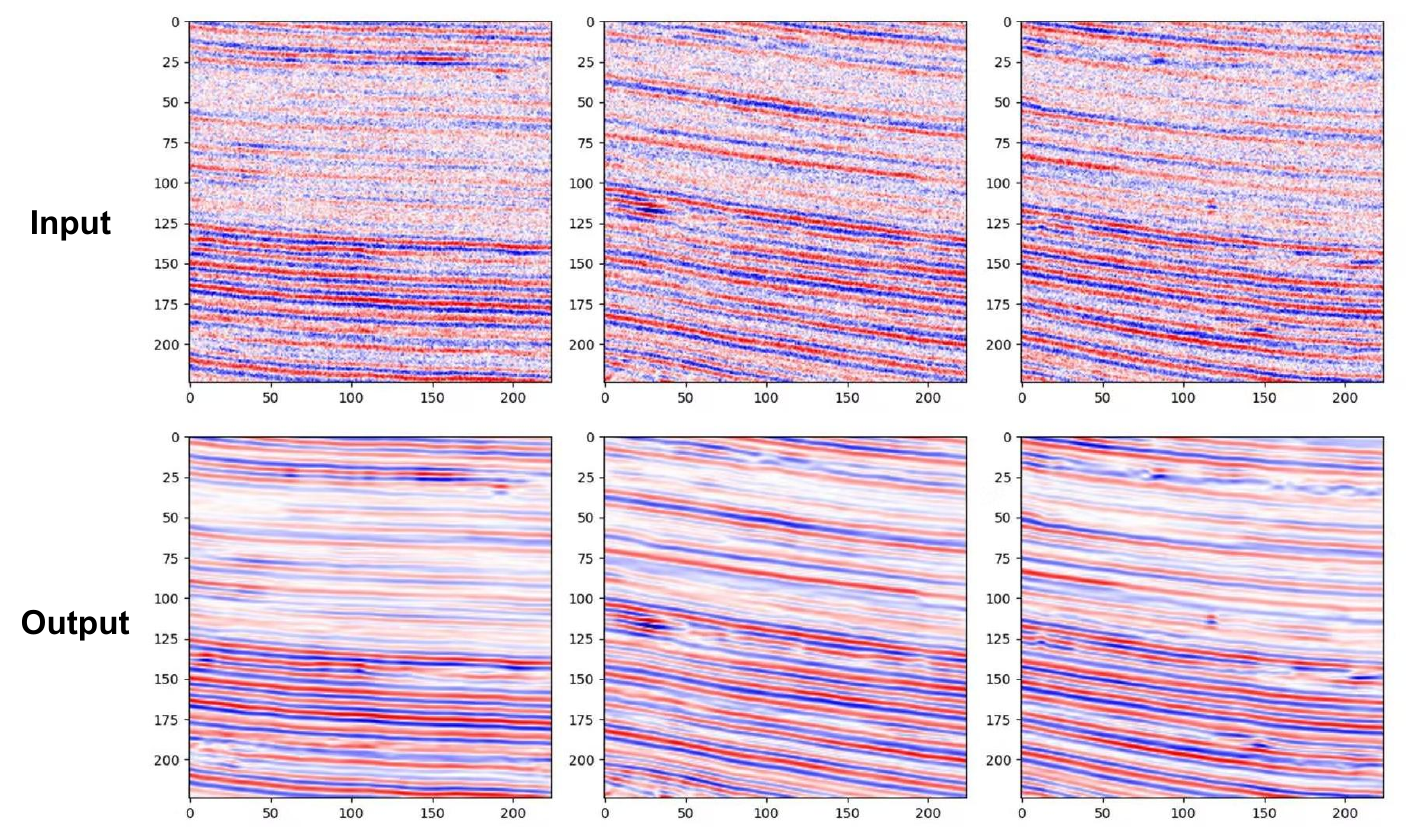} 
	\caption{The result diagram of SR-Agent when facing a single degraded seismic data with random noise.}
	\label{figure8}
\end{figure}

Fig.~\ref{figure8} shows the denoising results for seismic data contaminated by random noise. The degraded input contained obvious amplitude fluctuations and high-frequency disturbances, which obscured the continuity of seismic reflection events. After processing by the Transformer-based denoising network, the noise components were substantially suppressed while the major reflection structures were preserved.  The proposed denoising model achieved a PSNR of 34.6983 dB and an SSIM of 0.9942. The improvement in these metrics indicated that the specialized denoising network could effectively recover the underlying seismic signal from noisy observations. Importantly, the restoration result also exhibited improved spectral characteristics. The reduction of abnormal high-frequency energy and the preservation of dominant spectral components indicated that the denoising network did not simply suppress the signal amplitude, but learned to distinguish noise from useful seismic information.

 \begin{figure}
	\includegraphics[width=\linewidth]{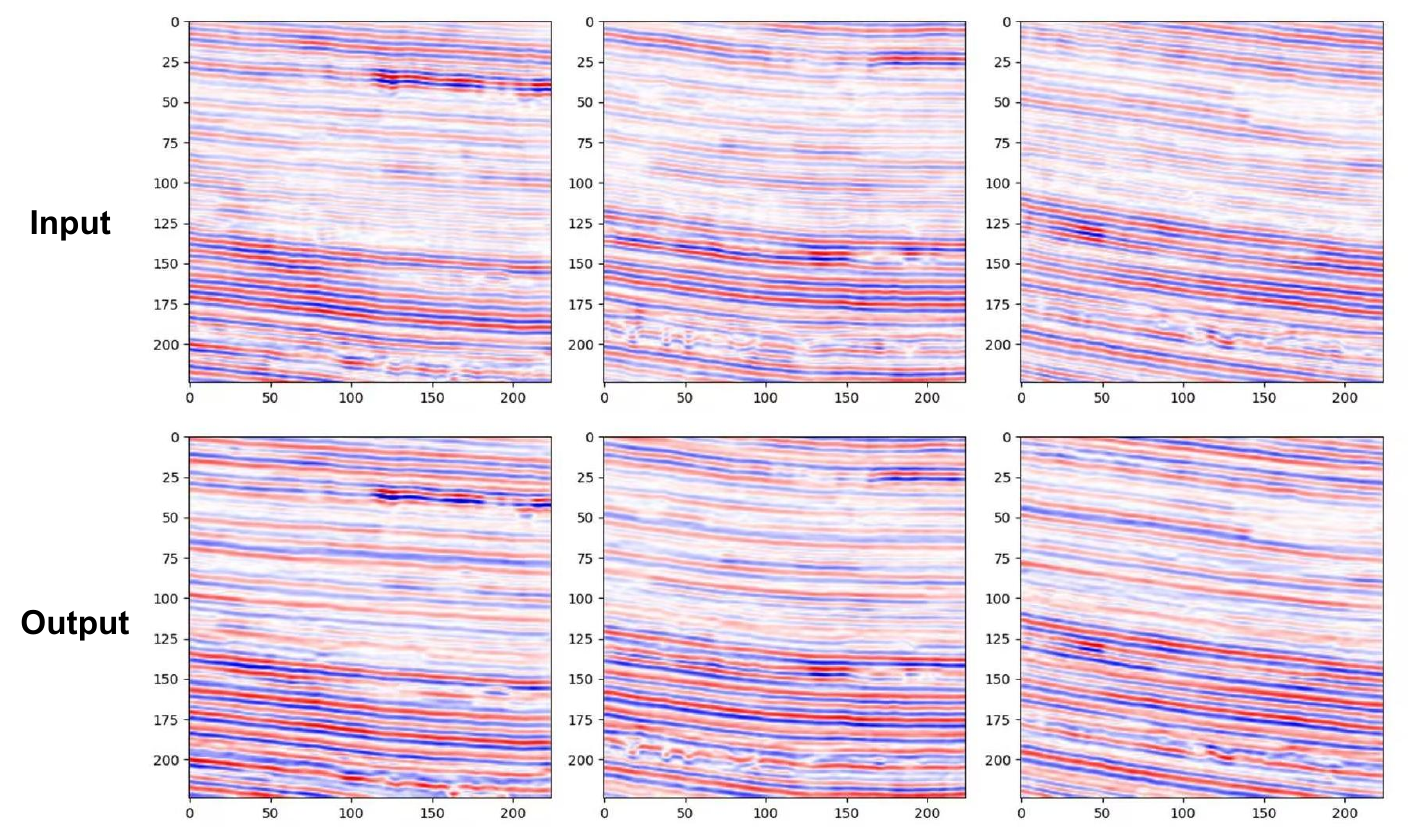} 
	\caption{The result diagram of SR-Agent when facing a single degraded seismic data with low-frequency loss.}
	\label{figure9}
\end{figure}

\begin{figure}
	\includegraphics[width=\linewidth]{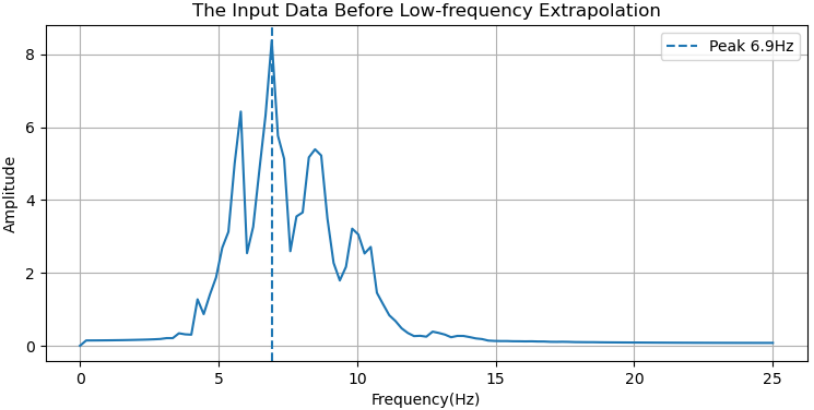} 
	\includegraphics[width=\linewidth]{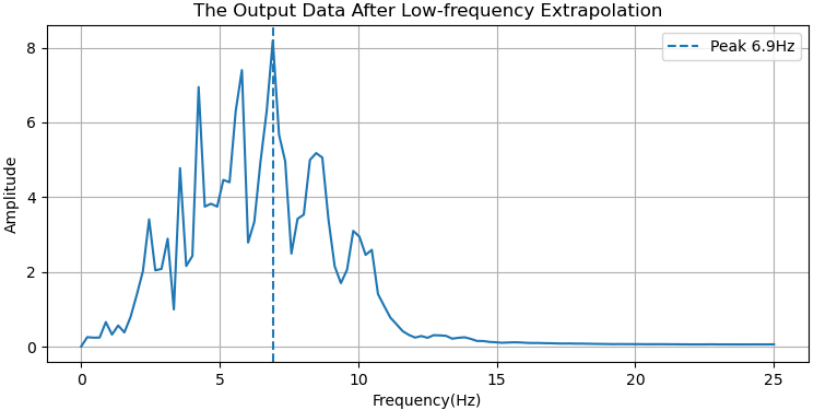} 
	\caption{The corresponding frequency spectra of Figure 9.The upper figure corresponds to the input data, and the lower figure corresponds to the output data.}
	\label{figure10}
\end{figure}

Fig.~\ref{figure9} shows the results of low-frequency extrapolation. The removal of low-frequency components caused a loss of the background spectral trend and may affect the continuity and amplitude characteristics of seismic events. The U-Net-based low-frequency extrapolation model restored the missing low-frequency components from the available higher-frequency information. The corresponding frequency spectra are shown in Fig.~\ref{figure10}. Compared with the degraded input, the restored data exhibited a broader frequency range and a spectral distribution closer to the reference data.

\begin{figure}
	\includegraphics[width=\linewidth]{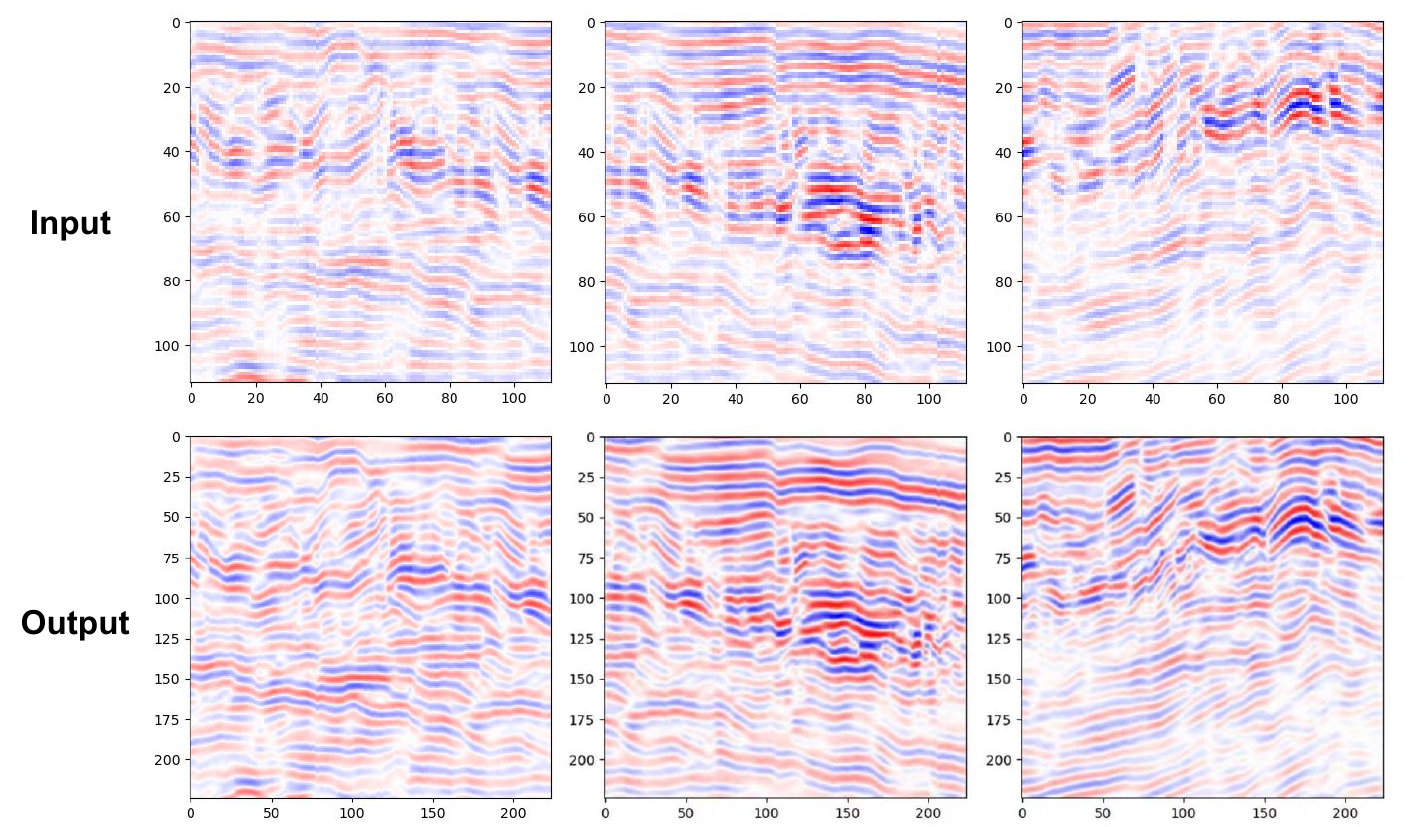} 
	\caption{The result diagram of SR-Agent when facing a single degraded seismic data with resolution enhancement.}
	\label{figure11}
\end{figure}

Fig.~\ref{figure11} shows the results of seismic resolution enhancement. The low-resolution input loss part of the spatial details because of the downsampling operation. The MAE-GAN-based resolution enhancement model made the reconstructed seismic events exhibit sharper spatial variations and clearer local structures.

Overall, the single-degradation experiments demonstrated that the four specialized models could effectively address their corresponding degradation types. These models therefore provided a reliable model pool for the subsequent agentic restoration experiments.

 \begin{figure}
	\includegraphics[width=\linewidth]{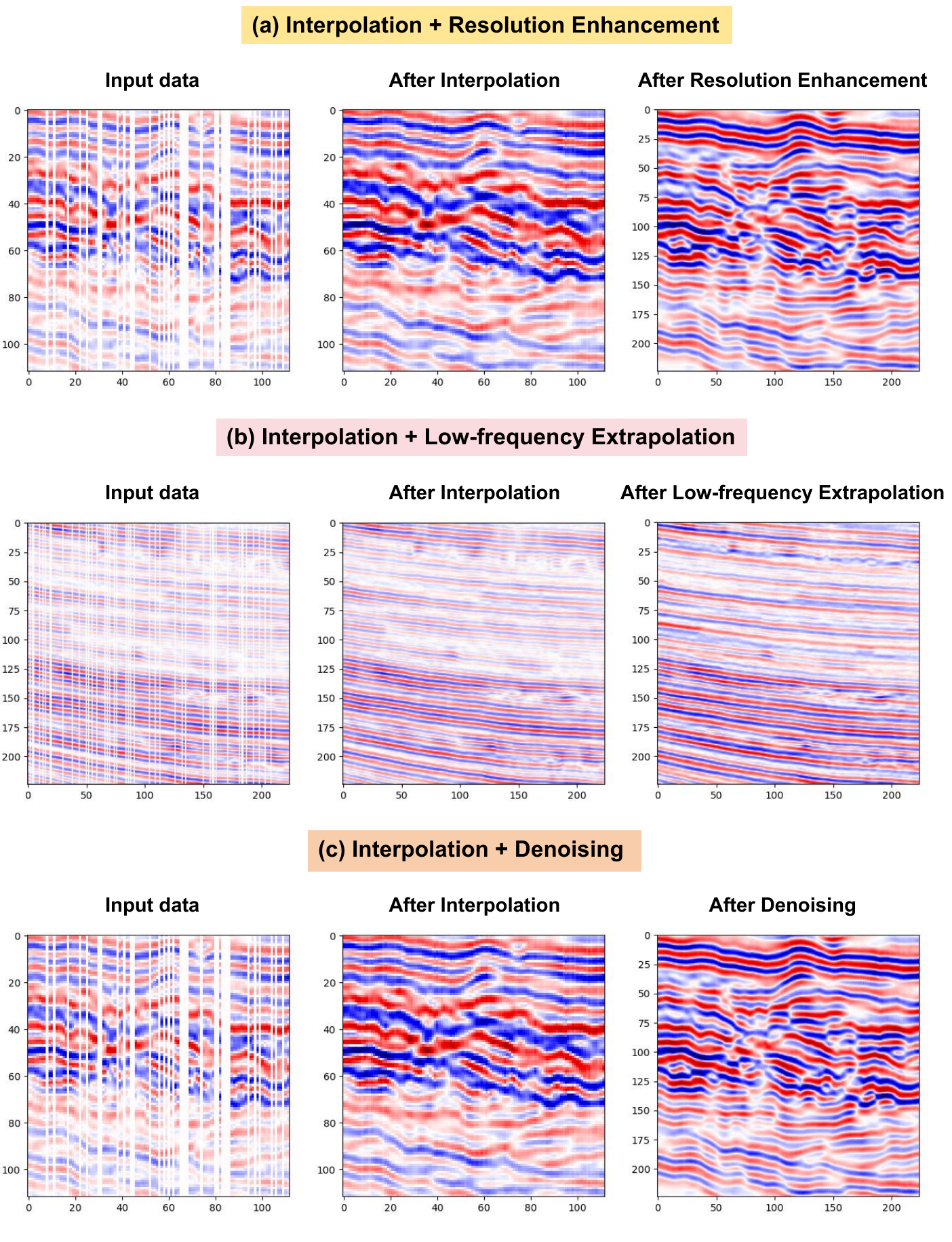} 
	\caption{The result diagram of SR-Agent when facing a single degraded seismic data with low-resolution.}
	\label{figure12}
\end{figure}

\subsection{Restoration under Multiple Degradations}
Although the individual restoration models demonstrate satisfactory performance for their respective tasks, real seismic data may simultaneously contain multiple degradation types. In such cases, the central challenge is no longer simply selecting an appropriate restoration model, but determining which models should be executed and in what order.
To investigate this problem, 3 dual-degradation cases are considered, including noise + missing traces, missing traces + low-frequency loss, and missing traces + resolution enhancement. For each multi-degradation input, SR-Agent first identifies the degradation types and subsequently generates candidate restoration sequences according to the detected degradation combination. Fig.~\ref{figure12} shows the representative results of the three dual-degradation cases.

 \begin{figure}
	\includegraphics[width=\linewidth]{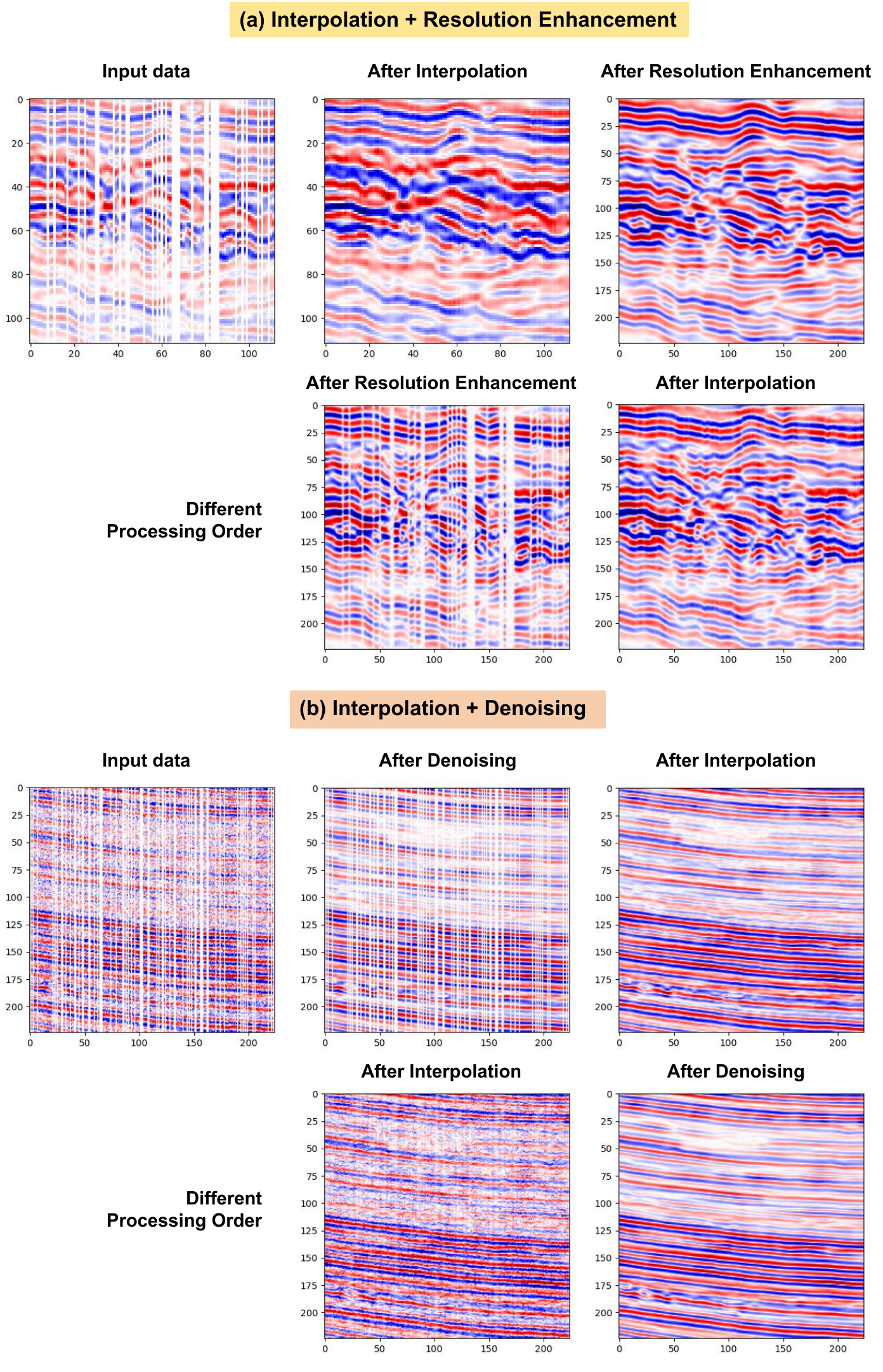} 
	\caption{The result diagram of SR-Agent when facing a single degraded seismic data with low-resolution.}
	\label{figure13}
\end{figure}

The processing order had a noticeable influence on the reconstruction results when multiple degradations coexisted. For the noise and missing-trace case, as shown in Fig.~\ref{figure13}(a), applying interpolation before denoising tended to result in the loss of some local seismic details. The regions highlighted by the boxes showed that several fine-scale waveform features became less distinct after the interpolation–denoising sequence. In contrast, performing denoising before interpolation better preserved these local features while providing a cleaner input for subsequent trace reconstruction.

A similar effect could be observed for the interpolation and resolution enhancement case in Fig.~\ref{figure13}(b). When resolution enhancement was performed before interpolation, unreasonable waveform fluctuations could be generated at the locations of missing traces. As highlighted by the boxes, these artificial fluctuations were inconsistent with the surrounding seismic events and became more apparent after subsequent interpolation. In comparison, performing interpolation first provided a more complete spatial sampling pattern, allowing the subsequent resolution enhancement model to enhance the seismic resolution with fewer spatial artifacts.

These observations demonstrated that the restoration operations were not independent. Instead, the output of one restoration model changed the input distribution of the next model. Therefore, treating multi-degradation restoration as a fixed sequential pipeline may lead to not good results.

\begin{figure*}
	\includegraphics[width=\textwidth]{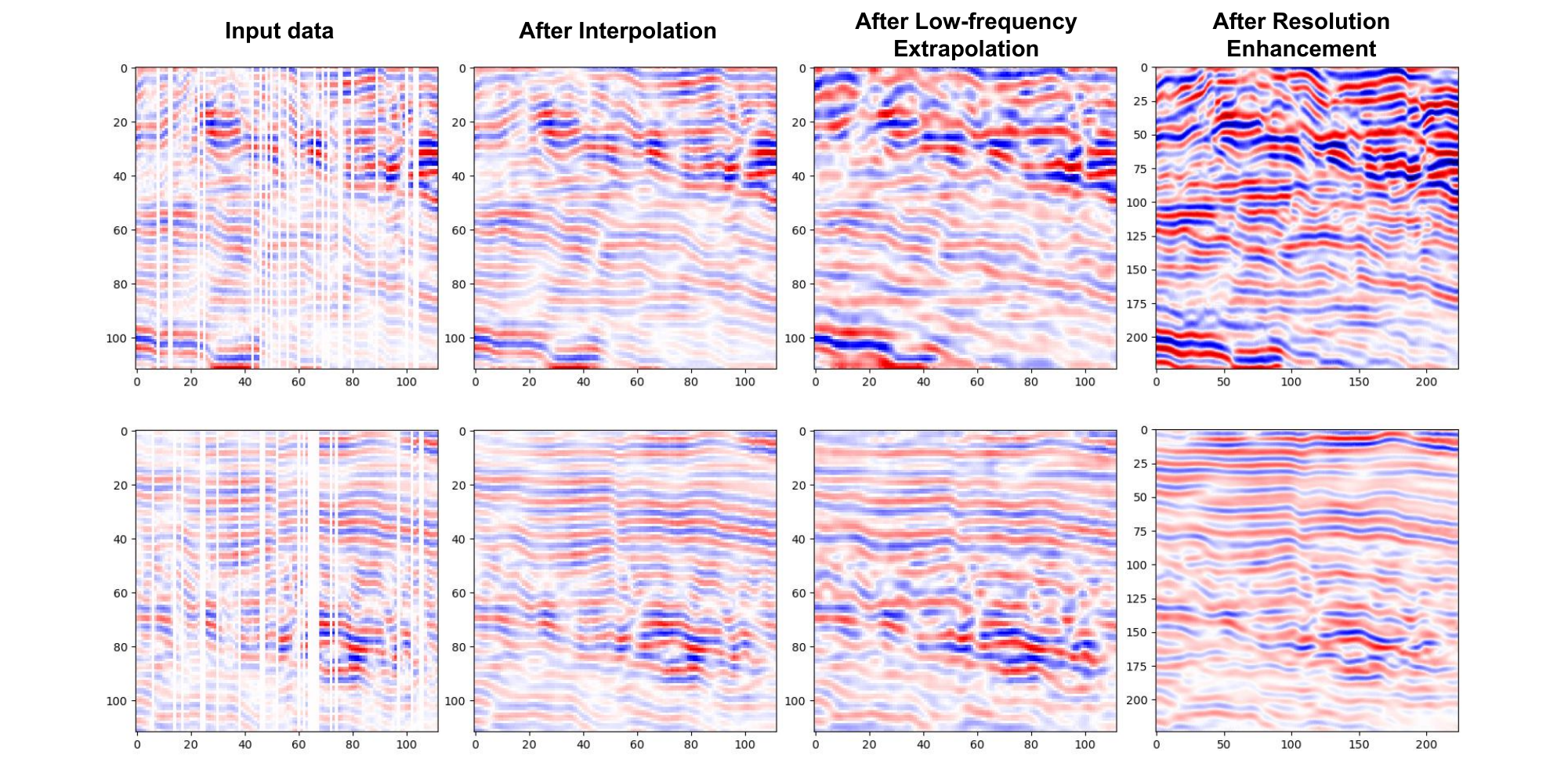} 
	\caption{The restoration results for missing traces + low-frequency loss + resolution enhancement.}
	\label{figure14}
\end{figure*}

To further evaluate the scheduling capability of SR-Agent, we additionally consider one king of three representative cases involving three simultaneous degradation types: missing traces + low-frequency loss + resolution enhancement. Multiple possible restoration sequences are generated from the detected degradation types, and the corresponding restoration results are evaluated using the proposed no-reference quality metric. The restoration results are shown in Fig.~\ref{figure14}. The results indicated that interpolation first provided a more complete spatial sampling pattern, after which low-frequency extrapolation and resolution enhancement could be performed with reduced spatial inconsistency. These results further demonstrated that the optimal restoration sequence depended on the specific combination and interaction of degradation types rather than following a universally fixed order.

Table 2 summarizes the quality scores obtained from different processing sequences for the three-degradation cases. Each candidate sequence corresponded to a different ordering of the three restoration operations, and the sequence with the highest quality score was selected by SR-Agent as the preferred restoration path. The variations in scores among different sequences demonstrated that restoration order could substantially affect the final reconstruction quality.

	\begin{table*}
		\centering
		\caption{The quality scores, PSNR, SSIM obtained from different processing
			sequences for missing traces + low-frequency loss + resolution enhancement.}
		\begin{tabular}{|c|c|c|c|c|c|c|c|}
			\hline
			\multicolumn{2}{|c|}{Step1} & \multicolumn{2}{c|}{Step2} & \multicolumn{4}{c|}{Step3} \\
			\hline
			Name of step & Score & Name of step & Score & Name of step & PSNR & SSIM & Score \\
			\hline
			{Interpolation} & {91.4053} & Low-frequency Extrapolation & 91.4448 & Resolution Enhancement & 29.2894 & 0.9336 & 93.2336 \\
			& & Resolution Enhancement & 92.7153 & Low-frequency Extrapolation & 28.6087 & 0.9013 & 90.5292 \\
			\hline
			{Low-frequency Extrapolation} & {83.3561} & Interpolation & 90.0079 & Resolution Enhancement & 26.1141 & 0.8587 & 92.8070 \\
			& & Resolution Enhancement & 89.2854 & Interpolation & 25.0647 & 0.8605 & 89.8881 \\
			\hline
			{Resolution Enhancement} & {84.4532} & Interpolation & 90.3228 & Low-frequency Extrapolation & 23.9098 & 0.8354 & 87.7921 \\
			& & Low-frequency Extrapolation & 81.8268 & Interpolation & 24.6411 & 0.8449 & 87.1251 \\
			\hline
		\end{tabular}
	\end{table*}
	
\subsection{Overall Results}

The experimental results demonstrated three main findings.

First, the proposed degradation classification network could effectively identify different degradation patterns by jointly exploiting spatial and frequency-domain information. This provided reliable perception capability for the subsequent agentic restoration process.

Second, the independently trained task-specific restoration networks provided effective restoration for their corresponding degradation types. Their specialization enabled each network to focus on the structural or spectral characteristics associated with a specific restoration task rather than simultaneously modeling multiple heterogeneous degradation mechanisms.

Third, SR-Agent could coordinate these specialized models through no-reference quality feedback. Compared with fixed restoration pipelines and all-in-one restoration strategies, the proposed framework achieved more flexible processing of multi-degradation seismic data. In particular, the results demonstrated that restoration order had a significant influence on the final reconstruction quality, and that evaluating alternative restoration paths could effectively reduce the risk of applying an unsuitable processing sequence.

Overall, these experiments validated the feasibility of treating seismic data restoration as an agentic decision-making problem rather than a conventional single-network reconstruction problem. The proposed SR-Agent therefore provided a framework for integrating heterogeneous seismic restoration models into a unified, adaptive, and closed-loop processing system.

\section{DISCUSsion}

\subsection{Discussion on Agentic Seismic Restoration}

Conventional seismic restoration methods generally treat restoration as a task-specific image-to-image mapping, in which a dedicated model is designed for a predefined degradation. Although such methods can achieve high accuracy under controlled conditions, their flexibility is limited when multiple degradation mechanisms occur simultaneously. In contrast, SR-Agent formulates seismic restoration as an agentic decision-making process consisting of degradation perception, restoration planning, model execution, and quality-guided feedback. Rather than requiring a single model to handle heterogeneous degradation patterns, SR-Agent employs specialized restoration models and dynamically determines their application according to the degradation condition.

The experimental results indicate that the main advantage of SR-Agent does not simply arise from the individual restoration networks, but from its ability to determine which model should be applied and in what order. Different restoration operations may interact with each other; for example, denoising before interpolation can reduce noise propagation, whereas interpolation before resolution enhancement can provide more complete spatial information. Therefore, a fixed restoration sequence may not always produce optimal results. By evaluating restoration quality after different operations, SR-Agent can adaptively select restoration paths and avoid unnecessary processing. Compared with all-in-one restoration paradigms, this modular strategy provides more explicit degradation recognition and better interpretability, although it requires additional computational cost for path evaluation. Such a trade-off is acceptable for offline seismic processing, while efficient scheduling or learned policies could further reduce the computational burden.

Another important component of SR-Agent is the no-reference quality assessment mechanism, which enables restoration decisions without requiring clean ground-truth seismic data. By evaluating structural continuity, FK-domain characteristics, spectral stability, and high-frequency energy, SR-Agent can monitor the quality of intermediate restoration results and determine whether further processing is necessary. This transforms the restoration process from a fixed sequential pipeline into a closed-loop framework and reduces the risk of excessive restoration and artifact accumulation. However, the current quality assessment relies mainly on manually designed seismic attributes and may not fully capture subtle geological structures or interpretation-oriented requirements. Future work could therefore explore learned quality assessment models based on expert knowledge, self-supervised learning, or foundation models to improve the adaptability and reliability of agentic restoration.

Overall, SR-Agent provides a flexible framework that shifts seismic restoration from single-task reconstruction toward perception-driven, quality-guided, and adaptive restoration. Its effectiveness is therefore determined not only by the capability of individual restoration models, but also by the intelligence of the decision-making and feedback mechanisms that coordinate them.

\subsection{Limitations and Future Perspectives}

Although SR-Agent provides a new framework for multi-degradation seismic restoration, several limitations remain.

First, the performance of SR-Agent degrades for certain complex combinations of seismic degradations, particularly when multiple restoration operations need to be performed sequentially. For example, combinations such as denoising followed by low-frequency extrapolation, or denoising, interpolation, and low-frequency extrapolation performed successively, may result in lower restoration quality than relatively simple degradation combinations. This limitation is mainly attributed to error accumulation during the sequential restoration process. Specifically, the output of an upstream restoration model is used as the input for subsequent models, and residual errors or distortions introduced during an earlier restoration step may be propagated and amplified by subsequent operations. As the number of sequential restoration steps increases, such accumulated errors may gradually affect seismic amplitude, phase, spectral characteristics, and structural continuity. Future studies could therefore investigate error-aware restoration strategies that explicitly monitor, suppress, or compensate for error propagation between successive restoration models. Joint optimization of multiple restoration operations or feedback-based correction mechanisms may also provide an effective way to alleviate error accumulation and improve the robustness of restoration for complex degradation combinations.

Second, although the proposed no-reference quality assessment module provides an effective quantitative basis for iterative restoration and path selection, its evaluation capability remains inherently limited by the characteristics of the adopted indicators. The four metrics, including seismic trace coherence, FK-domain energy distribution, spectral smoothness, and high-frequency energy ratio, mainly characterize the spatial continuity and frequency-domain properties of the restored seismic data. They therefore provide an objective description of whether the restoration result exhibits statistically and physically reasonable signal characteristics, but do not directly measure its deviation from the unknown true seismic response. This limitation is particularly important in practical seismic processing, where high-quality ground-truth data are generally unavailable. Consequently, an increase in the comprehensive quality score Q should be interpreted as an improvement in the measured signal characteristics rather than definitive evidence that the restored data are closer to the true subsurface response. In addition, the current quantitative assessment mainly focuses on signal-level characteristics and does not explicitly incorporate several properties that are critical for subsequent seismic interpretation and reservoir characterization, such as amplitude preservation, phase consistency, relative impedance information, and geological plausibility. A restoration result may achieve a higher quantitative score by producing smoother and more continuous seismic events while inadvertently altering subtle geological features or suppressing weak but meaningful reflections. Such effects may not be adequately reflected by the current no-reference metrics. Therefore, the quantitative evaluation should be considered complementary to, rather than a replacement for, geological interpretation and application-oriented validation.

Future work could address these limitations by developing a more comprehensive quality assessment framework that jointly considers signal fidelity, structural consistency, amplitude and phase preservation, and geological interpretability. In particular, physics-informed constraints and domain knowledge from seismic interpretation and reservoir characterization could be incorporated into the quality assessment and decision-making processes. Where partial reference data or reliable geological information are available, hybrid evaluation strategies combining no-reference metrics with reference-based and task-oriented measures could also provide a more robust assessment of restoration quality. Such developments would allow SR-Agent to distinguish not only whether a restoration result is quantitatively improved, but also whether the improvement corresponds to the preservation and recovery of geologically meaningful seismic information.

\section{conclusion}

We proposed SR-Agent, an agentic framework for multi-degradation seismic data restoration. The framework integrates degradation perception, restoration scheduling, task-specific model execution, and no-reference quality feedback into a closed-loop workflow. A spatial-frequency dual-branch classifier identifies degradation types, while specialized models for denoising, interpolation, low-frequency extrapolation, and resolution enhancement. Experiments show that SR-Agent can effectively coordinate specialized restoration models and adaptively select restoration sequences for multi-degradation seismic data. The proposed approach provides a basis for more autonomous seismic processing systems. Future work would focus on reducing error accumulation in longer restoration sequences and developing more comprehensive quality assessment methods that incorporate amplitude, phase, and geological interpretability.

\end{document}